# Relativistic free-motion time-of-arrival

**Zhi-Yong Wang***, **Cai-Dong Xiong**

School of Physical Electronics, University of Electronic Science and Technology of China, Chengdu 610054, CHINA

E-mail: zywang@uestc.edu.cn

**Abstract**

Relativistic free-motion time-of-arrival theory for massive spin-1/2 particles is systematically developed. Contrary to the nonrelativistic time-of-arrival operator studied thoroughly in previous literatures, the relativistic time-of-arrival operator possesses self-adjoint extensions because of the particle-antiparticle symmetry. The nonrelativistic limit of our theory is in agreement with the nonrelativistic time-of-arrival theory.

## 1. Introduction

In the traditional formalism of quantum theory, time enters as a parameter rather than a dynamical operator. As a consequence, the investigations on tunneling time, arrival time and traversal time, etc., still remain controversial today [1-19]. On one hand, one imposes self-adjointness as a requirement for any observable; on the other hand, according to Pauli's argument [20-23], there is no self-adjoint time operator canonically conjugating to a Hamiltonian if the Hamiltonian spectrum is bounded from below. A way out of this dilemma is based on the use of positive operator valued measures (POVMs) [19, 22-26]: quantum observables are generally positive operator valued measures, e.g., quantum observables are extended to maximally symmetric but not necessarily self-adjoint operators [15, 27-30], in such a way one preserves the requirement that time operator be conjugate to the Hamiltonian but abandons the self-adjointness of time operator.

However, all mentioned above are mainly based on the framework of nonrelativistic quantum mechanics. In this paper, arrival time is studied at the level of relativistic quantum mechanics, for the moment Pauli's objection is no longer valid. Historically, the first attempt was made to study a relativistic time-of-arrival can be found in Ref. [31], where via the Newton-Wigner position operator of the Klein-Gordon particle, the author introduced an operator for the time-of-arrival of the Klein-Gordon particle. Another later study relevant to relativistic time-of-arrival was given by A. Ruschhaupt [32], where, by applying the relativistic extension of Event-Enhanced Quantum Theory (which main idea is to view the total system as consisting of coupled classical and quantum part), the author has computed the relativistic time-of-arrival of a free particle with spin-1/2. In contrast with these works, our work is based on standard relativistic quantum mechanics of spin-1/2 particles (with nonzero mass), and lays emphasis on a directly relativistic extension for the traditional theory of nonrelativistic time-of-arrival. In the following, the natural units of measurement ($\hbar = c = 1$) is applied, repeated indices must be summed according to the Einstein rule, and the space-time metric tensor is chosen as $g^{\mu\nu} = \text{diag}(1,-1,-1,-1)$, $\mu,\nu = 0,1,2,3$.

**2. Relativistic free-motion time-of-arrival operator**

Let $\boldsymbol{\alpha} = (\alpha_1, \alpha_2, \alpha_3)$ denote a matrix vector, where $\alpha_i = \beta\gamma^i$ ($i = 1,2,3$), $\beta = \gamma^0$, and $\gamma^\mu$'s ($\mu = 0,1,2,3$) are the 4×4 Dirac matrices satisfying the algebra $\gamma^\mu\gamma^\nu + \gamma^\nu\gamma^\mu = 2g^{\mu\nu}$. A free spin-1/2 particle of rest mass $m$ has the Hamiltonian $\hat{H} = \boldsymbol{\alpha} \cdot \hat{\boldsymbol{p}} + \beta m$. For simplicity, we choose a coordinate system with its x-axis being parallel to the momentum of the particle, such that the four-dimensional (4D) momentum of the particle is $p^\mu = (E, p, 0, 0)$ (for our purpose, we assume that whenever $p \neq 0$, i.e., $E^2 \neq m^2$, this condition presents no problem for our issues.), the Hamiltonian becomes $\hat{H} = \alpha_1 \hat{p} + \beta m$, where $\hat{p} = -i\partial/\partial x$, and the Dirac equation becomes ($\hbar = c = 1$)

$$i\partial \psi(t,x)/\partial t = (\alpha_1 \hat{p} + \beta m)\psi(t,x). \qquad (1)$$

Here, from $\hat{H} = \boldsymbol{\alpha} \cdot \hat{\boldsymbol{p}} + \beta m$ to $\hat{H} = \alpha_1 \hat{p} + \beta m$, it is just a matter of choosing a coordinate system. Therefore, Eq. (1) as a special case of the usual Dirac equation, describes the 3+1 fermions associated with the representation of the (3,1) Clifford algebra, rather than the 1+1 fermions associated with the representation of the (1,1) Clifford algebra. In other words, in physics, a spin-1/2 particle cannot be related to the (1,1) Clifford algebra.

In order to study a time operator canonically conjugating to the Hamiltonian $\hat{H} = \alpha_1 \hat{p} + \beta m$, let us firstly introduce the common eigenstates of $\hat{H}$, $\hat{p}$ and the helicity operator, and denote them as $|p, \lambda, s\rangle$ in the momentum representation, while $|E, s\rangle$ in the energy representation. Where $|p, \lambda, s\rangle$'s satisfy the following orthonormality and completeness relations (owing to $\int_{-\infty}^{0} + \int_{0}^{+\infty} = \int_{-\infty}^{+\infty}$, the condition $p \neq 0$ has no effect on momentum integral)

$$\begin{cases} \langle p', \lambda', s'|p, \lambda, s\rangle = \delta_{\lambda\lambda'}\delta_{ss'}\delta(p-p') \\ \sum_{\lambda, s} \int_{-\infty}^{+\infty} |p, \lambda, s\rangle\langle p, \lambda, s| dp = I_{4\times 4} \end{cases} \qquad (2)$$

where $I_{4\times 4}$ is the 4×4 unit matrix (and so on), $p, p' \in (-\infty, 0) \cup (0, +\infty)$, $\lambda, \lambda' = \pm 1$ and $s, s' = \pm 1/2$. Let $|x\rangle$ and $|p\rangle$ respectively denote the position and momentum eigenstates, they satisfy $\langle x|p\rangle = \exp(ipx)/(2\pi)^{1/2}$. One can prove that $|p, \lambda, s\rangle = \varphi_{\lambda s}(p)|p\rangle$, where

$$\varphi_{\lambda s}(p) = \sqrt{\frac{m + \lambda E_p}{2\lambda E_p}} \begin{pmatrix} \eta_s \\ \dfrac{\sigma_1 p}{m + \lambda E_p} \eta_s \end{pmatrix}, \qquad (3)$$

where $E_p = \sqrt{p^2 + m^2}$, $\sigma_1$ is the x-component of the Pauli matrix-vector, and the two-component spinors $\eta_s$'s satisfy the orthonormality and completeness relations: $\eta_s^+\eta_{s'} = \delta_{ss'}$, $\sum_s \eta_s \eta_s^+ = I_{2\times 2}$, $\eta_s^+$ represents the hermitian conjugate of $\eta_s$ (and so on). In fact, the elementary solutions of Eq. (1) are $\psi_{p\lambda s}(t,x) = \langle x|p,\lambda,s\rangle \exp(-i\lambda E_p t)$. Let

$$|E,s\rangle \equiv [E^2/(E^2-m^2)]^{1/4}|p,\lambda,s\rangle, \qquad (4)$$

where $E = \lambda E_p \in \mathcal{R}_m \equiv (-\infty,-m)\cup(m,+\infty)$. In terms of $|E,s\rangle$ the orthonormality and completeness relations (2) can be rewritten as

$$\begin{cases} \langle E',s'|E,s\rangle = \delta_{ss'}\delta(E-E') \\ \sum_s \int_{\mathcal{R}_m} |E,s\rangle\langle E,s|\,dE = I_{4\times 4} \end{cases} \qquad (5)$$

Because

$$\hat{H}|E,s\rangle = E|E,s\rangle, \quad E \in \mathcal{R}_m = (-\infty,-m)\cup(m,+\infty), \qquad (6)$$

the Hamiltonian spectrum is $\mathcal{R}_m = (-\infty,-m)\cup(m,+\infty)$. In fact, Eqs. (2) and (5) show that, without the negative-energy part, the completeness requirement cannot be met and then the general solution of the Dirac equation cannot be constructed.

Now, let us introduce a time operator canonically conjugating to the Hamiltonian $\hat{H} = \alpha_1 \hat{p} + \beta m$. A natural way of introducing time operator is based on the usual quantization procedure. The classical expression for the arrival time at the origin $x_0 = 0$ of the freely moving particle having position $x$ and uniform velocity $v$, is $t = -x/v$. In the relativistic case, it is $t = -x/v = -x(E/p)$, where $E$ is the relativistic energy of the particle satisfying $E^2 = p^2 + m^2$. Replacing all dynamical variables with the corresponding linear operators, and symmetrizing the classical expression $t = -Ex/p$, one can obtain the relativistic time-of-arrival operator as follows (notice that $\hat{H}$ and $\hat{p}^{-1}$ commute such that a totally symmetrization is not necessary):

$$\hat{T} = -(1/4)[\hat{H}(\hat{p}^{-1}\hat{x} + \hat{x}\hat{p}^{-1}) + (\hat{p}^{-1}\hat{x} + \hat{x}\hat{p}^{-1})\hat{H}], \tag{7}$$

In the momentum representation, Eq. (7) becomes

$$\hat{T} = -\frac{i}{4}[H(p)(\frac{1}{p}\frac{\partial}{\partial p} + \frac{\partial}{\partial p}\frac{1}{p}) + (\frac{1}{p}\frac{\partial}{\partial p} + \frac{\partial}{\partial p}\frac{1}{p})H(p)]. \tag{8}$$

Inserting $\hat{H} = \alpha_1 \hat{p} + \beta m$ (or) $H(p) = \alpha_1 p + \beta m$ into Eq. (7) (or Eq. (8)), one can obtain the time-of-arrival operator of the free Dirac particle, say, $\hat{T}_{Dirac} = \hat{T}_{Dirac}(\hat{x}, \hat{p})$. It is easy to examine the canonical commutation relation $[\hat{T}_{Dirac}, \hat{H}] = -i$. Furthermore, applying Eqs. (2)-(5) and the relation $dE = pdp/E$, one can prove the following relation

$$\sum_{\lambda,s}\int_{-\infty}^{+\infty} dp \langle p, \lambda, s | \hat{T}_{Dirac}(\hat{x}, \hat{p}) | p, \lambda, s \rangle = \sum_{s}\int_{\mathcal{R}_m} dE \langle E, s | \hat{T}_{Dirac}(E) | E, s \rangle, \tag{9}$$

where

$$\hat{T}_{Dirac}(E) = -i\partial/\partial E. \tag{10}$$

Therefore, in the energy representation, the time-of-arrival operator is $-i\partial/\partial E$. In fact, the conclusion that an energy-representational time operator (not only the time-of-arrival operator) is $-i\partial/\partial E$ (or $i\partial/\partial E$, it is just a matter of convention), can be also found in the previous literatures [15, 28, 33-37].

**3. Eigenvalues and eigenfunctions of the relativistic time-of-arrival operator**

By inserting $\hat{H} = \alpha_1 \hat{p} + \beta m$ into Eq. (7) we get, in the position representation,

$$\hat{T}_{Dirac}(\hat{x}, \hat{p}) = -(\alpha_1 \hat{x} + \beta \hat{\tau}), \tag{11}$$

where

$$-\hat{\tau} = \hat{T}_{non}(\hat{x}, \hat{p}) = -m(\hat{p}^{-1}\hat{x} + \hat{x}\hat{p}^{-1})/2 \tag{12}$$

is the nonrelativistic time-of-arrival operator that has been studied thoroughly in previous literatures [11, 19, 21, 22, 36], and can be called the *proper* time-of-arrival operator. In fact, using $t = -xE/p$ one has $t^2 - x^2 = (\pm xm/p)^2 = (\pm \tau)^2$, and then the nonrelativistic time-of-arrival $-\tau = -xm/p$ plays the role of proper time-of-arrival. Correspondingly, the nonrelativistic time-of-arrival operator plays the role of proper time-of-arrival operator.

In the momentum representation, Eq. (11) becomes

$$\hat{T}_{\text{Dirac}}(\hat{x}, \hat{p}) = \frac{1}{p}(\alpha_1 p + \beta m)(-i\frac{\partial}{\partial p}) + i\beta \frac{m}{2p^2}. \tag{13}$$

Assume that its momentum-representational eigenequation is

$$\hat{T}_{\text{Dirac}}(\hat{x}, \hat{p})\phi(p) = t\phi(p). \tag{14}$$

Firstly, let us tentatively assume that $\phi(p) \sim \exp(i\lambda E_p t)$, one can obtain eigenfunctions of $\hat{T}_{\text{Dirac}}(\hat{x}, \hat{p})$ as follows:

$$\phi_{t\lambda s}(p) = [p^2/(p^2 + m^2)]^{1/4} \varphi_{\lambda s}(p) \exp(i\lambda E_p t)/(2\pi)^{1/2}, \tag{15}$$

where $\varphi_{\lambda s}(p)$ is given by Eq. (3). However, the exact value of the eigenvalue $t$ remains to be determined. For this, let us assume that $\phi(p) \sim \exp(-ipx)$, one can prove that the eigenvalues and eigenfunctions of $\hat{T}_{\text{Dirac}}(\hat{x}, \hat{p})$ can be expressed as, respectively:

$$\begin{cases} t = -(E/p)x = -(\lambda E_p/p)x \\ \phi_{x\lambda s}(p) = [p^2/(p^2 + m^2)]^{1/4} \varphi_{\lambda s}(p) \exp(-ipx)/(2\pi)^{1/2} \end{cases}. \tag{16}$$

That is, the eigenvalue $t = -xE/p$ corresponds to the classical expression of relativistic time-of-arrival, just as one expected. On the other hand, substituting the proper time-of-arrival $-\tau = -xm/p$ and the eigenvalues $t = -xE/p$ into Eq. (14) and solving it again, one can prove that the eigenvalues and eigenfunctions of $\hat{T}_{\text{Dirac}}(\hat{x}, \hat{p})$ can be also expressed as, respectively:

$$\begin{cases} t = -bt_x \\ \phi_{xbs}(p) = [x^2/(x^2 + \tau^2)]^{1/4} \xi_{bs}(x) \exp(-ipx)/(2\pi)^{1/2} \end{cases}, \tag{17}$$

where $b = \pm 1$, $t_x = \sqrt{x^2 + \tau^2}$, and

$$\xi_{bs}(x) = \sqrt{\frac{\tau + bt_x}{2bt_x}} \begin{pmatrix} \eta_s \\ \dfrac{\sigma_1 x}{\tau + bt_x} \eta_s \end{pmatrix}. \tag{18}$$

As we know, within the propagator theory, Dirac antiparticles can be interpreted as particles of negative energy moving backwards in space and time [38-41], and then related to the fact that there are both positive and negative energy solutions, there are both positive and negative time-of-arrivals, and they describe the time-of-arrivals of particles and antiparticles, respectively.

Consider that the eigenfunctions $\phi_{x\lambda s}(p) \equiv \langle p | x, \lambda, s \rangle$ (or $\phi_{xbs}(p) \equiv \langle p | x, b, s \rangle$) correspond to the momentum representation of the eigenstates $|x, \lambda, s\rangle$ (or $|x, b, s\rangle$), using $\langle p | x \rangle = \exp(-ipx)/(2\pi)^{1/2}$ and Eqs. (17)-(18), one has

$$\begin{cases} |x, \lambda, s\rangle = [p^2/(p^2 + m^2)]^{1/4} \varphi_{\lambda s}(p) |x\rangle \\ |x, b, s\rangle = [x^2/(x^2 + \tau^2)]^{1/4} \xi_{bs}(x) |x\rangle \end{cases} . \tag{19}$$

Contrary to the nonrelativistic case, using Eq. (19) one can show that the eigenstates of $\hat{T}_{\text{Dirac}}$ form an orthogonal and complete set, e.g.,

$$\begin{cases} \langle x', \lambda', s' | x, \lambda, s \rangle = [p^2/(p^2 + m^2)]^{1/2} \delta_{\lambda\lambda'} \delta_{ss'} \delta(x - x') \\ \sum_{\lambda s} \int dx |x, \lambda, s\rangle\langle x, \lambda, s| = [p^2/(p^2 + m^2)]^{1/2} I_{4\times 4} \end{cases} . \tag{20}$$

The time-of-arrival operator $\hat{T}_{\text{Dirac}} = -(\alpha_1 \hat{x} + \beta \hat{\tau})$ is related to the position operator $\hat{x}$, as a result, via Eq. (19) the eigenstates of $\hat{T}_{\text{Dirac}}$ are related to those of $\hat{x}$, such that their spatial behaviors (including the locality) are similar to those of $|x\rangle$. For example, in the position representation, the eigenfunctions of $\hat{T}_{\text{Dirac}}$ satisfy $\langle x' | x, b, s \rangle \sim \delta(x' - x)$. In particular, as $m = 0$ (or $\tau = 0$), one has $\hat{T}_{\text{Dirac}} = -\alpha_1 \hat{x}$, and Eq. (17) becomes

$$\begin{cases} t = \mp x \\ \phi_{xbs}(p) = \dfrac{1}{\sqrt{2}} \begin{pmatrix} \eta_s \\ \pm \sigma_1 \eta_s \end{pmatrix} \exp(-\mathrm{i}px)/(2\pi)^{1/2} \end{cases} \quad (21)$$

Eq. (21) shows that, in the momentum representation, excepting the spin wavefunction $\dfrac{1}{\sqrt{2}} \begin{pmatrix} \eta_s \\ \pm \sigma_1 \eta_s \end{pmatrix}$ (being a 4×1 constant matrix), $\phi_{xbs}(p)$'s are the momentum-representational eigenfunctions of the position operator $\hat{x}$, just as one expected. From the point of view of classical mechanics, as $m=0$ (or $\tau=0$), along the direction of motion space is equivalent to time.

**4. Self-adjoint extensions of the relativistic time-of-arrival operator**

Consider that some terminologies in different literatures have different meanings, or their meanings in physics are different from those in mathematics, to avoid confusing, let us unify the definitions for linear operator mapping the Hilbert space $\mathcal{H}$ into itself as follows: 1) The operator $\hat{F}$ is Hermitian if $\langle \psi | \hat{F} \varphi \rangle = \langle \hat{F} \psi | \varphi \rangle$, $\forall \psi, \varphi \in D(\hat{F})$, $\bar{D}(\hat{F}) \subset \mathcal{H}$, where $D(\hat{F})$ is the domain of $\hat{F}$, $\bar{D}(\hat{F})$ is the closed set of $D(\hat{F})$; 2) The operator $\hat{F}$ is symmetric if $\langle \psi | \hat{F} \varphi \rangle = \langle \hat{F} \psi | \varphi \rangle$, $\forall \psi, \varphi \in D(\hat{F})$, $\bar{D}(\hat{F}) = \mathcal{H}$; 3) The operator $\hat{F}$ is self-adjoint if it is symmetric and $\hat{F}^+ = \hat{F}$, $D(\hat{F}^+) = D(\hat{F})$, so that $\langle \psi | \hat{F} \varphi \rangle = \langle \hat{F}^+ \psi | \varphi \rangle$; 4) The operator $\hat{F}$ is essentially self-adjoint if it is symmetric and has exactly one self-adjoint extension. It possesses self-adjoint extensions if and only if its deficiency indices are equal.

Firstly, Eqs. (11)-(12) show that the singularity of $\hat{T}_{\text{Dirac}}(\hat{x}, \hat{p})$ is the same as that of the nonrelativistic time-of-arrival operator $\hat{T}_{\text{non}}(\hat{x}, \hat{p})$, the latter has been studied in Ref. [19, 22]. Therefore, our results here are similar to those in Ref. [19, 22]: $\mathcal{D}(\hat{T}_{\text{Dirac}})$ is the set of absolutely continuous square integrable functions of p on the real line, and $\|\hat{T}_{\text{Dirac}}\varphi\|$ is finite. Therefore, the singularity of $\hat{T}_{\text{Dirac}}$ at $p=0$ is avoided. An alternative way out of this singularity can be found in Ref. [28, 29, 33, 34, 42], where time operator is represented by a bilinear operator.

As shown by Eqs. (9)-(10), in the energy representation, the time operator becomes $\hat{T}_{\text{Dirac}}(E) = -i\partial/\partial E$, where $E \in \mathcal{R}_m = (-\infty, -m) \cup (m, +\infty)$, and then its domain $\mathcal{D}(\hat{T}_{\text{Dirac}})$ can be taken as a dense domain of the Hilbert space of square integrable functions on $\mathcal{R}_m = (-\infty, -m) \cup (m, +\infty)$, which is a subspace of square integrable absolutely continuous functions (say, $\varphi(E)$) whose derivative is also square integrable provided that $\varphi(\pm m) = 0$. Using $\varphi(\pm m) = 0$ one can easily show that $\hat{T}_{\text{Dirac}}(E) = -i\partial/\partial E$ is symmetric.

Further, because the Hamiltonian spectrum is $\mathcal{R}_m = (-\infty, -m) \cup (m, +\infty)$, the deficiency indices of $\hat{T}_{\text{Dirac}}$ satisfy $n_+ = n_-$, where $n_\pm = \dim \text{Ker}(\hat{T}^+_{\text{Dirac}} \mp iI)$, $I$ denotes an identity operator, $\text{Ker}(\hat{F}) \equiv \{\varphi \in \mathcal{H} \mid \hat{F}\varphi = 0)\}$ is the kernel of $\hat{F}$, and $\dim(S)$ denotes the dimension of the space $S$. Therefore, $\hat{T}_{\text{Dirac}}$ has self-adjoint extension. However, in the present paper, it is difficulty for us to ascertain whether $\hat{T}_{\text{Dirac}}$ has exactly one self-adjoint extension (i.e., whether $\hat{T}_{\text{Dirac}}$ is an essentially self-adjoint operator), this is not the purpose of the paper. Obviously, as $m = 0$, $\hat{T}_{\text{Dirac}}$ is a self-adjoint operator, which can be also shown from another point of view: as $m = 0$, $\hat{T}_{\text{Dirac}} = -\alpha_1 \hat{x}$, where $\hat{x}$ belongs to the position space while $\alpha_1$ belongs to the Dirac-spinor space, they are separately self-adjoint and satisfy $\alpha_1 \hat{x} = \hat{x}\alpha_1$, then $\hat{T}_{\text{Dirac}}$ is self-adjoint.

As we know, the coexistence of the positive- and negative-energy solutions is associated with particle-antiparticle symmetry, where antiparticles can be interpreted as particles of negative energy moving backwards in space and time [38-41]. Eqs. (2) and (5) show that, without the positive- or negative-energy part, the completeness requirement cannot be met and then the general solution of the Dirac equation cannot be constructed. For example, to obtain a wave-packet with Gaussian density distribution, a superposition of plane waves of positive as well as of negative energy is necessary [43]. Moreover, in relativistic quantum mechanics, observables are characterized by the probability distributions of measurement results in both positive- and negative-energy states, and the probability distributions for the relativistic time-of-arrival can be influenced by the interference between the positive- and negative-energy compounds of a wave-packet. Therefore, in our case, the negative-energy solution cannot be discarded such that the time operator $\hat{T}_{\text{Dirac}}$ has self-adjoint extensions.

## 5. Nonrelativistic limit

Now, let us study the nonrelativistic limit of the eigenvalues and eigenfunctions of the relativistic time-of-arrival operator $\hat{T}_{\text{Dirac}}$. Using $E_p^2 - m^2 = p^2$ let us rewrite Eq. (3) as the usual form:

$$u(p,s) = \varphi_{+s}(p) = \sqrt{\frac{m+E_p}{2E_p}} \begin{pmatrix} \eta_s \\ \frac{\sigma_1 p}{m+E_p} \eta_s \end{pmatrix}, \tag{22}$$

$$w(p,s) = \frac{\sigma_1 p}{|p|} \varphi_{-s}(-p) = \sqrt{\frac{m+E_p}{2E_p}} \begin{pmatrix} \frac{\sigma_1 p}{m+E_p} \eta_s \\ \eta_s \end{pmatrix}. \tag{23}$$

In the nonrelativistic limit, one has

$$u(p,s) \to \begin{pmatrix} \eta_s \\ 0 \end{pmatrix} \equiv \zeta_{+s}(p), \quad w(p,s) \to \begin{pmatrix} 0 \\ \eta_s \end{pmatrix} \equiv \zeta_{-s}(p). \tag{24}$$

That is, the nonrelativistic limit of $\varphi_{\lambda s}(p)$ is equal to $\zeta_{\lambda s}(p)$ ($\lambda = \pm 1$). As we know, the general solution of the Dirac equation is a four-component spinor (say, 4-spinor). Eq. (24) shows that, in the nonrelativistic limit, the positive-energy solution ($\lambda = 1$) alone forms the upper 2-spinor of the 4-spinor, while the negative-energy solution ($\lambda = -1$) alone forms the lower 2-spinor of the 4-spinor. For the moment, the general solution of the Dirac equation no longer contains a coherent superposition between the positive- and negative-energy components, and in terms of 2-spinors, one can shows that the completeness relation can be satisfied alone by the positive- or negative-energy solution (for the moment the completeness relation concerns the unit matrix $I_{2\times 2}$ rather than $I_{4\times 4}$). Therefore, in the nonrelativistic limit, one can separately analyze the positive-energy and the negative-energy components. Consider that antiparticles can be interpreted as particles of negative energy moving backwards in space and time, when we separately study the positive- and negative-energy components, the corresponding Hamiltonian spectrum can be taken as $(m, +\infty)$. Therefore, in the nonrelativistic limit, we will only consider the positive-energy solution, i.e., take $E = E_p = \sqrt{p^2 + m^2}$ only.

Firstly, the nonrelativistic limit of the eigenvalue of $\hat{T}_{\text{Dirac}}$ is

$$t = -xE/p \rightarrow t_{\text{non}} = -xm/p, \tag{25}$$

where $t_{\text{non}} = -xm/p$ is the eigenvalue of the nonrelativistic time-of-arrival operator $\hat{T}_{\text{non}}$.

In order to compare our results with those presented in the traditional theory of nonrelativistic time-of-arrival, let us study the nonrelativistic limit of Eq. (15) with $\lambda = 1$, and from now on we omit the subscript $\lambda$. To do this we split the time dependence of $\phi_{t\lambda s}(p) = \phi_{ts}(p)$ into two terms, that is, in the nonrelativistic limit, let

$$\exp(\mathrm{i}Et) = \exp[\mathrm{i}\,p^2 t/2m)\exp(\mathrm{i}mt)\,, \tag{26}$$

where the term containing the kinetic energy represents the nonrelativistic time-evolution factor, and then in the nonrelativistic limit, the term containing the rest mass should be omitted. Therefore, the nonrelativistic limit of Eq. (15) is

$$\phi_{ts}(p) \to \phi_{\mathrm{non}ts}(p) = (p^2/m^2)^{1/4}\,\zeta_s(p)\exp(\mathrm{i}\,p^2 t/2m)/(2\pi)^{1/2}\,. \tag{27}$$

Except for the 4-spinor $\zeta_{\lambda s}(p) = \zeta_s(p)$ that stands for the spin wave-function, the remainder of $\phi_{\mathrm{non}ts}(p)$ is just the eigenfunction of the nonrelativistic time-of-arrival operator $\hat{T}_{\mathrm{non}}$, which due to the fact that, the traditional theory of nonrelativistic time-of-arrival takes no account of particle's spin.

## 6. Time operator: further considerations

It is interesting to note that, the time operator $\hat{T}_{\mathrm{Dirac}} = -\alpha_1 \hat{x} - \beta \hat{\tau}$ is to $t^2 = x^2 + \tau^2$ as the Hamiltonian $\hat{H} = \alpha_1 \hat{p} + \beta m$ is to $E^2 = p^2 + m^2$, which shows us a duality between the position and momentum space. Because of $[\hat{x}, \hat{T}_{\mathrm{Dirac}}] \neq 0$, there is an uncertainty relation between the time-of-arrival and position-of-arrival. Consider that the time operator is $-\mathrm{i}\partial/\partial E$ in the energy representation, one can formally introduce a dual counterpart of the Schrödinger equation $\mathrm{i}\partial \psi(t)/\partial t = \hat{H}\psi(t)$, namely

$$-\mathrm{i}\frac{\partial}{\partial E}\phi(E) = \hat{T}\phi(E)\,. \tag{28}$$

According to Ref. [44], one can call $\hat{T}$ "time-Hamiltonian", or, seeing that a Hamiltonian can be called energy function, one can also call $\hat{T}$ "time function" [37]. However, to provide Eq. (28) with physical meanings, the energy $E$ and the momentum $p$ contained in $\phi(E)$ should be regarded as two independent variables (that is, $\partial E/\partial p = 0$), and $\phi(E)$ represents an "event state" satisfying $t^2 = x^2 + \tau^2$ rather than a "particle state" satisfying the $E^2 = p^2 + m^2$. As $\hat{T} = \hat{T}_{\text{Dirac}}(\hat{x}, \hat{p})$, using Eq. (19) one can prove that the elementary solutions of Eq. (28) can be expressed as $\phi_{xbs}(E, p) = \langle p | x, b, s \rangle \exp(ibt_x E)$, namely

$$\phi_{xbs}(E, p) = [x^2/(x^2 + \tau^2)]^{1/4} \xi_{bs}(x) \exp[i(tE - xp)]/(2\pi)^{1/2}, \qquad (29)$$

where $t = bt_x = b\sqrt{x^2 + \tau^2}$. As mentioned before, the elementary solutions of Eq. (1) can be expressed as $\psi_{p\lambda s}(t, x) = \langle x | p, \lambda, s \rangle \exp(-i\lambda E_p t)$, and then we can regard $\phi_{xbs}(E, p)$ as a dual counterpart of $\psi_{p\lambda s}(t, x)$.

Therefore, under the dual transformation of $x^\mu = (t, x) \leftrightarrow p^\mu = (E, p)$, one has the following dual relations: Eq. (3)$\leftrightarrow$Eq. (18), and

$$\begin{cases} t^2 = x^2 + \tau^2 \leftrightarrow E^2 = p^2 + m^2 \\ \hat{T} = -\alpha_1 \hat{x} - \beta \hat{\tau} \leftrightarrow \hat{H} = \alpha_1 \hat{p} + \beta m \\ -i\partial\phi(E)/\partial E = \hat{T}\phi(E) \leftrightarrow i\partial\psi(t)/\partial t = \hat{H}\psi(t) \\ \phi(E) \sim \phi_{xbs}(E, p) \leftrightarrow \psi(t) \sim \psi_{p\lambda s}(t, x) \end{cases}. \qquad (30)$$

It is important to note that, as for Eq. (28) describing the event state $\phi(E)$, in which $E$ and $p$ are taken as two independent variables ($\partial E/\partial p = 0$) while $t$ and $x$ not (owing to $t^2 = x^2 + \tau^2$); conversely, as for Eq. (1) describing the particle state $\psi(t)$, in which $t$ and $x$ are two independent variables ($\partial x/\partial t = 0$) while $E$ and $p$ not (owing to $E^2 = p^2 + m^2$). A completely dual approach can be found in Ref. [45].

## 7. Conclusions

Up to now, the theory of time-of-arrival is extended from nonrelativistic to relativistic quantum-mechanical case, where the eigenvalues and eigenfunctions of the relativistic time-of-arrival operator are given. Due to the particle-antiparticle symmetry, the relativistic time-of-arrival operator possesses self-adjoint extensions, which also in agreement with the fact that, in order to obtain relativistic quantum mechanics, space and time have to be treated equally. As for a free Dirac particle, its time-of-arrival operator $\hat{T}_{\text{Dirac}} = -\alpha_1 \hat{x} - \beta \hat{\tau}$ is to $t^2 = x^2 + \tau^2$, as its Hamiltonian operator $\hat{H} = \alpha_1 \hat{p} + \beta m$ is to $E^2 = p^2 + m^2$, which displays a duality between coordinate space and momentum space. A correct nonrelativistic limit of our theory is obtained.

**Acknowledgment**

The first author (Z. Y. Wang) would like to greatly thank Prof. A. Ruschhaupt and Prof. J. G. Muga for their helpful discussions and valuable suggestions. Project supported by the Specialized Research Fund for the Doctoral Program of Higher Education of China (Grant No. 20050614022) and by the National Natural Science Foundation of China (Grant No. 60671030).